\documentclass[twocolumn]{article}
\usepackage[a4paper, total={180mm, 243mm}]{geometry}
\usepackage[utf8]{inputenc}
\usepackage{amssymb}
\usepackage{lipsum}
\usepackage{graphicx}
\usepackage{siunitx}
\usepackage[version=4]{mhchem} 
\usepackage{titling}
\usepackage{circledsteps}   
\usepackage{subcaption}
\usepackage{soul}
\usepackage[switch]{lineno} 
\modulolinenumbers[5]

\DeclareCaptionLabelSeparator{custom}{\textbar}
\DeclareCaptionFormat{custom}{\textsf{\textbf{#1 #2} \small #3}}
\newcommand\authormark[1]{\textsuperscript{#1}}

\graphicspath{{./figures}} 

\makeatletter
\newcommand{\showfontsize}{\f@size{} pt}
\makeatother

\usepackage[
    backend=biber,
    style=nature,
    url=false,
    isbn=false,
    eprint=true,
    date=year
]{biblatex}
\newcommand{\citet}[1]{\cite{#1}}
\newcommand{\citep}[1]{\cite{#1}}
\newcommand{\citealt}[1]{\cite{#1}}

\DeclareRobustCommand{\ion}[2]{%
  \relax
  \ifmmode
    \ifx\testbx\f@series
      {\mathbf{#1\,\mathsc{#2}}}
    \else
      {\mathrm{#1\,\mathsc{#2}}}
    \fi
  \else
    \textup{#1\,{\mdseries\textsc{#2}}}%
  \fi
 }

\definecolor{Myblue}{cmyk}{0.8,0.6,0,0.1}
\newcommand{\rev}[1]{{{#1}}}

\title{Ultraviolet astronomical spectrograph calibration with \\laser frequency combs from nanophotonic \rev{lithium niobate} waveguides}

\author{Markus Ludwig,\authormark{1,*}
        Furkan Ayhan,\authormark{2,*}
        Tobias M. Schmidt,\authormark{3,*} \\
        Thibault Wildi,\authormark{1} 
        Thibault Voumard,\authormark{1} 
        Roman Blum,\authormark{4}
        Zhichao Ye,\authormark{5}   
        Fuchuan Lei,\authormark{5} \\
        François Wildi,\authormark{3} 
        Francesco Pepe,\authormark{3} 
        Mahmoud A. Gaafar,\authormark{1}  
        Ewelina Obrzud,\authormark{4} 
        Davide Grassani,\authormark{4} \\
        Olivia Hefti,\authormark{4} 
        Sylvain Karlen,\authormark{4} 
        Steve Lecomte,\authormark{4}
        François Moreau,\authormark{6}
        Bruno Chazelas,\authormark{3}
        Rico Sottile,\authormark{6} \\
        Victor Torres-Company,\authormark{5}
        Victor Brasch,\authormark{7} 
        Luis G. Villanueva,\authormark{2}
        François Bouchy,\authormark{3}
        Tobias Herr\authormark{1,8,**}
}

\date{%
    \small $^1$Deutsches Elektronen-Synchrotron DESY, Notkestr. 85, 22607 Hamburg, Germany \\
    \small $^2$École Polytechnique Fédérale de Lausanne (EPFL), 1015 Lausanne, Switzerland \\
    \small $^3$ Observatoire de Gen\`eve, Département d’Astronomie, Universit\'e de Gen\`eve, Chemin Pegasi 51b, 1290 Versoix, Switzerland \\
    \small $^4$ Swiss Center for Electronics and Microtechnology (CSEM), 2000 Neuch\^atel, Switzerland\\
    \small $^5$ Department of Microtechnology and Nanoscience, Chalmers University of Technology, 41296 Gothenburg, Sweden \\
    \small $^6$ Observatoire de Haute-Provence, CNRS, Université d'Aix-Marseille, 04870 Saint-Michel-l'Observatoire, France \\
    \small $^7$ Q.ANT GmbH, Handwerkstraße 29, 70565 Stuttgart, Germany\\
    \small $^8$Physics Department, Universität Hamburg UHH, Luruper Chaussee 149, 22607 Hamburg, Germany\\
    \small $^*$These authors contributed equally.\\
    \small $^{**}$tobias.herr@desy.de
}

\begin{document}

\maketitle


\textbf{
Astronomical precision spectroscopy underpins searches for life beyond Earth, direct observation of the expanding Universe and constraining the potential variability of physical constants across cosmological scales. Laser frequency combs can provide the critically required accurate and precise calibration to the astronomical spectrographs. For cosmological studies, extending the calibration with such \textit{astrocombs} to the ultraviolet spectral range is highly desirable, however, strong material dispersion and large spectral separation from the established infrared laser oscillators have made this exceedingly challenging. Here, we demonstrate for the first time astronomical spectrograph calibrations with an astrocomb in the ultraviolet spectral range below 400 nm. This is accomplished via chip-integrated highly nonlinear photonics in periodically-poled, nano-fabricated lithium niobate waveguides in conjunction with a robust infrared electro-optic comb generator, as well as a chip-integrated microresonator comb. These results demonstrate a viable route towards astronomical precision spectroscopy in the ultraviolet and may contribute to unlocking the full potential of next generation ground- and future space-based astronomical instruments.
}

\noindent
Precise astronomical spectroscopy has led to the Nobel-Prize winning discovery of the first exo-planet via the \textit{radial velocity method} \citep{mayor1995}, which relies on the precise tracking of minute Doppler-shifts in the stellar spectrum caused by an orbiting planet. The detection of Earth-like planets in the habitable zone around a Sun-like star requires to measure a radial velocity change as small as 10~cm/s (relative Doppler-frequency shift of $3\times10^{-10}$) over the time scale of 1~year.
This necessitates regular wavelength calibration of the spectrograph against a suitable calibration light source.
Highly stable and accurate calibration light sources hence take on a key role in astronomical precision spectroscopy.

\begin{figure}[t!]
  \centering
  \includegraphics[width=\columnwidth]{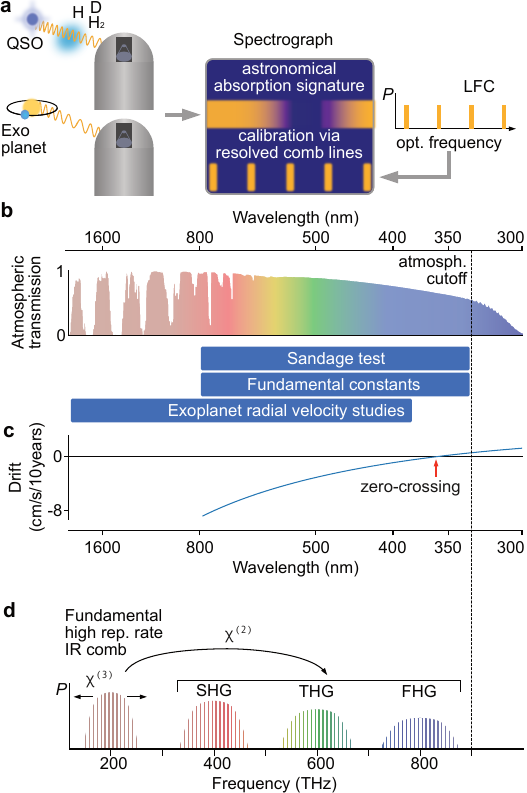}
  \caption{
    \textbf{Ultraviolet astronomical precision spectroscopy}.
        \textbf{a}, Precision spectroscopy of astronomical objects enabled by laser frequency combs (LFC) providing absolutely and precisely known laser frequencies, cf. main text. QSO: quasi stellar object; H,H$_2$: hydrogen; D: deuterium; \textit{P}: optical power.
        \textbf{b}, Atmospheric transmission (high altitude, dry), spectral ranges for different astrophysical studies. \textbf{c}, Expected redshift drift signal for the standard $\Lambda$CDM cosmology \cite{aghanim2020} in units of cm/s per decade (Sandage test) as function of wavelength, corresponding to the redhifted \ion{H}{I} Ly$\alpha$ transition.
        \textbf{d}, Concept of broadening and transferring a high-repetition rate astrocomb from infrared (IR) to visible (VIS) and ultraviolet (UV) wavelengths based on optical $\chi^{(2)}$- and $\chi^{(3)}$-nonlinearities.
        }
  \label{fig:intro}
\end{figure}

Calibration sources with outstanding accuracy and precision can be provided by laser frequency combs \cite{fortier2019, diddams2020} (LFCs). Their spectra are comprised of large sets of discrete laser lines whose frequencies are precisely and absolutely known (\rev{linked to a metrological frequency standard including the SI second}), on a level that exceeds the astronomical requirements.
Provided their lines can be separated by the spectrograph, LFCs can serve as exquisite wavelength calibrators \cite{murphy2007, fischer2016} and are then also referred to as \textit{astrocombs} (Fig.~\ref{fig:intro}a). Recent years have seen significant progress in developing astrocombs \citep{steinmetz2008, li2008, braje2008, mccracken2017, herr2019} and extending their coverage to visible (VIS) wavelengths, where the emission of Sun-like stars peaks, and to near-infrared wavelength, where relatively cold stars can accommodate rocky planets in tight orbits.

Extending astrocombs further towards the atmospheric cutoff and into the ultraviolet (UV) regime \rev{would not only be relevant to exo-planet science, but} highly desirable for fundamental physics and precision cosmology (Fig.~\ref{fig:intro}b);  \rev{indeed these} are also major science drivers for the upcoming ANDES high-resolution spectrograph at the \textit{Extremely Large Telescope} (ELT) \citep{marconi2022}. For instance, by accurately determining the wavelengths of narrow metal absorption features in the spectra of quasi-stellar objects (QSOs), one can infer the value of the fine-structure constant, $\alpha$, at the place and time of absorption, i.e. billions of years ago. This allows to test whether fundamental physical constants might have varied on cosmological scales \citep{webb1999, murphy2022}.  Another exciting, but extremely demanding science case is the direct observation of cosmic expansion in real time, often referred to as the \textit{Sandage test} \citep{sandage1965, liske2008, cristiani2023}. This requires detecting a drift of \ion{H}{I} Ly$\alpha$ forest absorption features in QSO spectra by just a few cm/s (relative frequency shift of $3\times10^{-11}$) over several decades. 
\rev{As lined out in} \cite{martins2023}, \rev{observations and corresponding wavelength calibrations in the UV are needed} to access low redshifts, $z$, in particular the important zero-crossing of the drift at $z=2$, corresponding to $365\,\mathrm{nm}$ (Fig.~\ref{fig:intro}c). \rev{Achieving such an extreme level of wavelength stability will require a community effort over the next decade to develop improved spectrographs, data analysis methods, and in particular LFC-based calibration sources.}
\rev{Furthermore, it has recently become clear that a detailed modeling of the instrumental line-spread function is crucial to further improve accuracy and stability of spectroscopic observations} \cite{hirano2020, milakovic2023, martins2023}. 
\rev{This, however, requires a broadband calibration source with a spectrum of narrow lines, such as those of an LFC.}

Astrocombs can be derived from high-repetition rate ($f_\mathrm{rep}>10$~GHz) laser sources, such as line-filtered mode-locked lasers \cite{steinmetz2008, li2008, braje2008, benedick2010}, electro-optic (EO) comb generators \cite{kashiwagi2016, yi2016a} and chip-integrated microresonators \cite{kippenberg2018, pasquazi2018, gaeta2019, suh2019, obrzud2019b}. Their spectra consist of equidistant optical laser frequencies $\nu_n=\nu_0 + n f_\mathrm{rep}$ , spaced by the laser's pulse repetition rate $f_\mathrm{rep}$ and with an offset frequency $\nu_0$, which we here define to be the central frequency of the laser ($n$ is an integer comb line index).
To transfer the comb spectra to the desired wavelength range and to increase their spectral coverage, second ($\chi^{(2)}$) and third order ($\chi^{(3)}$) nonlinear optical effects are utilized. However, the high repetition rates of $>10$~GHz, that are necessary for the comb lines to \rev{be separated} by the spectrograph, imply low pulse energies and hamper nonlinear optical effects. In addition, and especially at shorter wavelengths, strong material dispersion complicates the efficient nonlinear energy transfer. 

So far, short-wavelength blue and green astrocombs have been achieved via second harmonic generation of infrared mode-locked lasers \cite{benedick2010, wilken2012, phillips2012} and of EO combs \cite{metcalf2019a} in bulk $\chi^{(2)}$-nonlinear crystals.
Spectral broadening in $\chi^{(3)}$-nonlinear photonic crystal fiber resulted in astrocombs spanning from infrared (IR) to VIS wavelength \cite{glenday2015, mccracken2017a, wu2022}, some extending to below 450~nm. 
In addition to these approaches, micro- and nanophotonic waveguides made of highly nonlinear materials can further increase the efficiency in a compact setup. 
In the IR, $\chi^{(3)}$-nonlinear silicon nitride (Si$_3$N$_4$) waveguides have led to broadband astrocombs \cite{metcalf2019}, and via third harmonic generation a 10~GHz IR comb has been transferred to VIS wavelengths \cite{obrzud2019}. 
Strong nonlinear effects in waveguides, including harmonic frequency generation, may also be obtained via the quadratic $\chi^{(2)}$-nonlinearity. At low pulse repetition rates, spectra across multiple optical octaves have been obtained in large-mode area ($\sim$100~µm$^2$) periodically poled lithium niobate (LiNbO$_3$) waveguides with mm-long propagation distances \cite{phillips2011, iwakuni2016, hickstein2017, rutledge2021}. 
Notably, periodically poled LiNbO$_3$ waveguides were utilized with a 30~GHz repetition rate line-filtered erbium-doped mode-locked laser \cite{nakamura2023}, creating harmonic spectra in VIS and UV domains for astronomical spectrograph calibration. In addition, recent work leveraged a combination of second harmonic and sum-frequency generation based on a titanium-doped sapphire laser to obtain, in conjunction with line-filtering, a continuous green to ultraviolet 30~GHz astrocomb \cite{cheng2024ContinuousUltravioletBluegreena}. 
In nano-photonic waveguides from thin-film LiNbO$_3$\cite{desiatov2019, honardoost2020, qi2020, zhu2021, churaev2022, bres2023}, owing to their sub-µm$^2$ modal confinement, highly efficient UV spectral generation has been reported \cite{wang2018, chen2019, yu2019, lu2019, chen2019, jankowski2020, reigescale2020, okawachi2020, sayem2021, obrzud2021, park2022, wu2024}, including at 10 GHz repetition rate \cite{wu2024}. However, astronomical spectrograph calibration in the UV spectral range has so far not been demonstrated.

Here, we demonstrate astronomical spectrograph calibration with laser frequency combs in the UV below 400~nm. Specifically, we design chip-integrated nano-structured LiNbO$_3$ waveguides with sub-µm$^2$ mode cross-section and a tailored poling-pattern to enable efficient UV astrocomb generation via cascaded harmonic generation from a robust 
telecommunication-wavelength 18~GHz EO comb, as well as, a 25~GHz chip-based microresonator comb.
The generated astrocombs are tested in March 2023 on the high-resolution echelle spectrograph \mbox{SOPHIE} \cite{perruchot2008} at the Observatoire de Haute Provence (OHP, France), achieving radial-velocity precision in the range of 1 to 2~m\,s$^{-1}$.

\subsection*{Results}
We base our astrocombs on a robust, telecommunication-grade EO comb with a pulse repetition rate of 18~GHz, as well as, a microresonator comb with a repetition rate of 25~GHz. Both are derived from an IR continuous-wave (CW) laser of frequency $\nu_0 \approx 192$~THz (wavelength $\approx$ 1.56~µm) and operate intrinsically at high repetition rate, so that filtering of unwanted comb lines is not required. 
To transfer the IR comb to the UV (and VIS), we utilize chip-based LiNbO$_3$ nanophotonic waveguides, for cascaded harmonic generation as illustrated in Fig.~\ref{fig:intro}d. 
The possible comb frequencies are 
\begin{equation}
    \nu_{m,n} = m\nu_0 + nf_\mathrm{rep}
    \label{eq:comb_freqs}
\end{equation}
where $m$ is the harmonic order. In general, each harmonic $m$ of the comb spectrum has a different offset frequency $m \nu_0$. While this can facilitate detection of the comb's offset frequency \cite{yu2019, okawachi2020, obrzud2021, wu2024}, care must be taken to avoid inconsistent frequency combs from overlapping harmonics with different offsets. Alternatively, stabilizing $\nu_0$ to a multiple of $f_\mathrm{rep}$ could address this point. 
\rev{Practically this could be achieved for instance by locking the beatnote between adjacent harmonics to zero} \cite{okubo2018}\rev{, or via an auxiliary low-repetition rate frequency comb }\cite{obrzud2018}.

\rev{A key difference between broadband electro-optic combs obtained via conventional self-phase modulation $\chi^{(3)}$-based supercontinua and spectra generated via $\chi^{(2)}$-based harmonics, lies in the phase noise properties of their comb lines. In $\chi^{(3)}$-based spectra, the phase noise, impacting linewidth and line contrast, grows quadratically with the frequency distance from the fundamental pump laser. In contrast, in $\chi^{(2)}$-based harmonic spectra the phase noise (approximately) depends on the frequency distance from it's corresponding harmonic's center frequency (see Supplementary Information SI, Section~1). This property, along with the efficiency of the nanophotonic waveguides, permits us to directly generate comb lines in the ultraviolet from a telecom-wavelength electro-optic source, within a certain bandwidth around the harmonic's center frequency; future extension of this bandwidth will, as we discuss in detail in the Section~1 in the SI, require the addition of a noise filtering cavity, which has been established for broadband $\chi^{(3)}$-based electro-optic combs} \cite{beha2017}, \rev{but is not implemented in this initial proof-of-concept demonstration.}

\begin{figure*}[ht!]
  \centering
  \includegraphics[width=\textwidth]{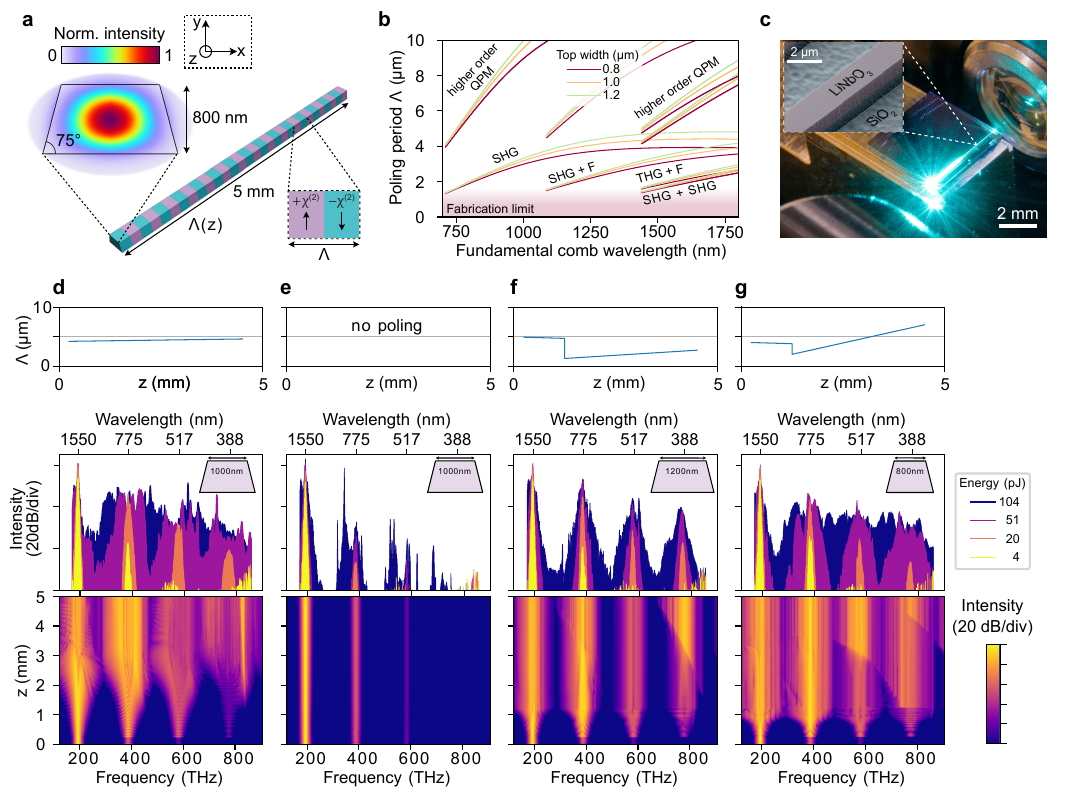}
  \caption{
    \textbf{Nanophotonic lithium niobate waveguides}.
        \textbf{a}, Geometry of a nanophotonic lithium niobate (LiNbO$_3$) waveguide, showing the optical mode profile for the fundamental transverse electric field (TE) mode and a possible poling pattern with spatially variable poling period $\Lambda$. The extraordinary crystal axis is oriented along the x-direction to access the highest electro-optic tensor element of LiNbO$_3$.
        \textbf{b}, Poling periods $\Lambda$ required for quasi-phase matching of various nonlinear optical processes for waveguides of different (top) width as a function of the fundamental comb's wavelength. SHG: second harmonic generation;  SHG + F: sum-frequency generation of second harmonic and fundamental comb; THG + F: sum-frequency generation of third harmonic and fundamental comb; SHG + SHG: sum-frequency generation of the second harmonic. Respective higher-order phase matching with three-fold period $\Lambda$ is indicated (the current fabrication limit only allows for $\Lambda>2$~µm).  
        \textbf{c}, Photograph of a waveguide in operation and scanning electron microscope image of the LiNbO$_3$ on silica (SiO$_2$). 
        \textbf{d,e,f,g}, Examples of waveguide designs showing poling pattern (top), experimentally generated spectra for different input pulse energies provided by a 100~MHz, 80~fs mode-locked laser with central wavelength of 1560~nm (\rev{waveguide cross-section are shown as insets and on chip-pulse energy is indicated by the color code}) and \textit{pyChi} \cite{voumard2023} simulation results (bottom) for a pulse energy of 50~pJ. \rev{The spurious spikes observed in the traces for 4~pJ pulse energy are manifestations of the noise floor of the optical spectrum analyzer (Yokogawa AQ6374).}
 }
  \label{fig:waveguideDesign}
\end{figure*}

\begin{figure*}[ht!]
  \centering
  \includegraphics[width=\textwidth]{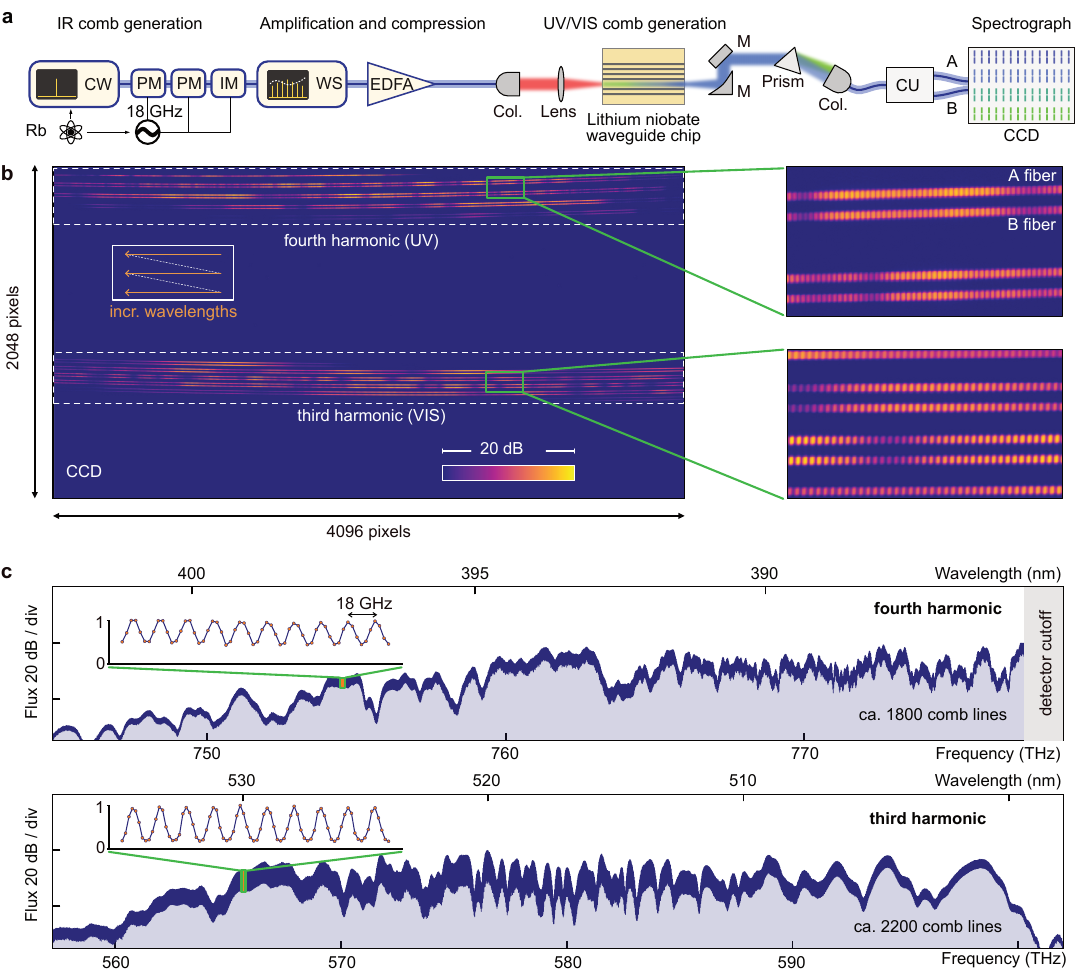}
  \caption{
    \textbf{Ultraviolet astrocombs}.
        \textbf{a}, Setup for ultraviolet (UV) astrocomb generation. Rb: rubidium-based atomic references (cf. main text, Methods); CW: continuous-wave laser; PM: phase modulator; IM: intensity modulator; WS: waveshaper; EDFA: erbium-doped fiber amplifier; Col.: fiber-to-free-space collimator; M: mirror; \rev{CU: calibration unit splitting the comb equally into A and B fibers;} CCD: SOPHIE charged coupled device detector. \rev{All fibers following the lithium niobate chip and feeding the spectrograph are multi-mode fibers.}
        \textbf{b}, Composite CCD detector image, obtained with two different settings of the flux-balancing prism,  
        showing fourth and third harmonic in the UV and visible (VIS) wavelength ranges. The inset indicates how wavelengths increase along the individual cross-dispersed echelle orders. The magnified CCD sub-images show the distinguishable but slightly overlapping frequency comb lines on the detector, originating from the two spectrograph fibers, which are both fed by the same comb. \textbf{c}, Extracted 1-dimensional comb spectra of both harmonics; the insets show a magnified (linear scale) portion of the spectrum and the 18~GHz-spaced comb lines. The UV detector cutoff is marked.
 }
  \label{fig:sophie}
\end{figure*}

\paragraph{Waveguide design.} An illustration of a nano-structured LiNbO$_3$ waveguide and the strong mode confinement to sub-µm$^2$ cross-section for efficient nonlinear conversion is shown in Figure~\ref{fig:waveguideDesign}a. 
The waveguides are fabricated from a commercially-available \rev{undoped} lithium niobate-on-insulator (LNOI) substrate with an 800~nm-thick LiNbO$_3$ layer via electron-beam lithography and reactive-ion etching, resulting in a sidewall angle of $\sim$75$^\circ$ (Fig.~\ref{fig:waveguideDesign}c, Methods). \rev{The waveguides taper to a width of 2500~nm at the chip facet, for input and coutput coupling.}
In a first design step, \rev{considering the fundamental transverse electric polarization (TE) waveguide mode}, we define the waveguide geometry, which \rev{in nano-photonic waveguides enables engineering of} the waveguide's group velocity dispersion (GVD), and \rev{additional} $\chi^{(3)}$-nonlinear spectral broadening of the input IR comb. In unpoled waveguides, top widths in the range of 800 to 1200~nm result in anomalous GVD and the most significant broadening. Spectral broadening is also possible in the normal GVD regime, typically resulting in less broad, but more uniform spectra. 


In a second design step, we define the waveguide's poling pattern \rev{to transfer the broadened fundamental spectrum to shorter wavelengths. This can be achieved efficiently}, when a fixed phase relation between the involved waves is maintained (\textit{phase matching}). However, due to chromatic material dispersion, this cannot easily be fulfilled \rev{across the entire spectral interval of interest (regardless of the choice of waveguide geometry and GVD at the fundamental wavelength)}. As an alternative \rev{to natural phase matching}, in ferroelectric crystals such as LiNbO$_3$, \textit{quasi-phase matching (QPM)} is possible. Here, the sign of the $\chi^{(2)}$-nonlinearity is periodically flipped within segments of length $\Lambda$ (poling period), via poling of the crystal's domains (Methods) to compensate the increasing phase-mismatch between the waves. For sum-frequency generation processes (which underlie harmonic generation), $\Lambda$ is
\begin{equation}
    \Lambda = 2\pi/| k_1 + k_2 - k_3 |
    \label{eq:qpm}
\end{equation}
where $k_1$ and $k_2$ are the input propagation constants, and $k_3$ is the output propagation constant in the waveguide. To achieve broadband QPM, $\Lambda$ may be chirped (i.e., varied along the waveguide).
To design $\Lambda(z)$ (z is the spatial coordinate in the propagation direction), we numerically compute the frequency $\omega$ dependent propagation constants $k(\omega)$ of the fundamental waveguide mode 
for waveguide top widths of 800, 1000 and 1200~nm. Based on Eq.~\ref{eq:qpm}, we derive the required poling period $\Lambda$ for a number of sum-frequency processes that would result in second harmonic generation (SHG = fundamental+fundamental), third harmonic generation (THG = SHG+fundamental) and fourth harmonic generation (FHG = SHG+SHG or THG+fundamental). These periods $\Lambda$ are shown in Fig.~\ref{fig:waveguideDesign}b, as a function of the fundamental comb's wavelengths. 

Prior to astrocomb generation, we test different waveguide designs with a low-repetition rate (100~MHz), 80~fs mode-locked laser operating at a center wavelength of 1560~nm. \rev{The pulses are coupled to the chip via a high-numerical aperture lens (NA=0.7) lens and the spectra are collected using reflective collimation optics and a fluoride multimode fiber that connects to a grating-based optical spectrum analyzer (Yokogawa AQ6374). Utilizing a lensed fiber with a calibrated coupling efficiency as temporary output coupler, we determine the input coupling efficiency of the high-NA lens to be $13\pm2$\%}. To illustrate the impact of the poling structure, we compare in Figure~\ref{fig:waveguideDesign}d a waveguide poled for broadband SHG ($\Lambda$ weakly chirped as indicated) with an unpoled waveguide (Fig.~\ref{fig:waveguideDesign}e), demonstrating markedly different behaviors. The poled waveguide creates a broadband SHG at approximately 400~THz, whereas the unpoled waveguide only creates narrow SHG. 
These observations agree with numerical simulation via \textit{pyChi} \cite{voumard2023} (shown below the experimental spectra). \rev{Note that the poled waveguide also exhibits a more substantial broadening of the input spectrum (and its harmonics), which can be explained through the contributions of cascaded $\chi^{(2)}$-processes to the effective $\chi^{(3)}$-nonlinearity.}
In a more advanced design (Fig.~\ref{fig:waveguideDesign}f) a short SHG section \rev{(supporting also broadening)} is followed by a longer and strongly chirped section designed to enable FHG. Indeed pronounced FHG is visible, although, in contrast to the simulation, it is not stronger than the THG signal; we attribute this to the very small values of $\Lambda<2$~µm, which are challenging to fabricate with high fidelity. 
To overcome the fabrication challenge for very small $\Lambda$, \rev{we also consider higher-order QPM} (Fig.~\ref{fig:waveguideDesign}b). 
In higher-order QPM, the poling period is an \rev{integer multiple $q$ of the fundamental poling period $\Lambda$} \cite{suhara1990}, \rev{which relaxes fabrication requirements at the cost of reduced efficiency. In our design for symmetric poling (equal length for both crystal orientations), $q$ is an odd-integer (here: 3); 
In addition, unintentional phase-matching may occur involving higher-order waveguide modes and coupling between orthorgonal polarizations.} 
A possible design is shown in Fig.~\ref{fig:waveguideDesign}g, where, after a short length of SHG poling,  $\Lambda$ ranges from 2 to 7~µm. This choice of $\Lambda$ respects the fabrication limit and provides QPM for most processes (Fig.~\ref{fig:waveguideDesign}b). The generated harmonic spectrum extends across 600~THz (ca. 350 to 1000~nm) within $\sim$20~dB of dynamic range, demonstrating, in agreement with previous work \cite{wu2024}, the potential of LiNbO$_3$ waveguides for generating gap-free spectra in the VIS and UV domains.

\begin{figure*}[ht!]
  \centering
  \includegraphics[width=\textwidth]{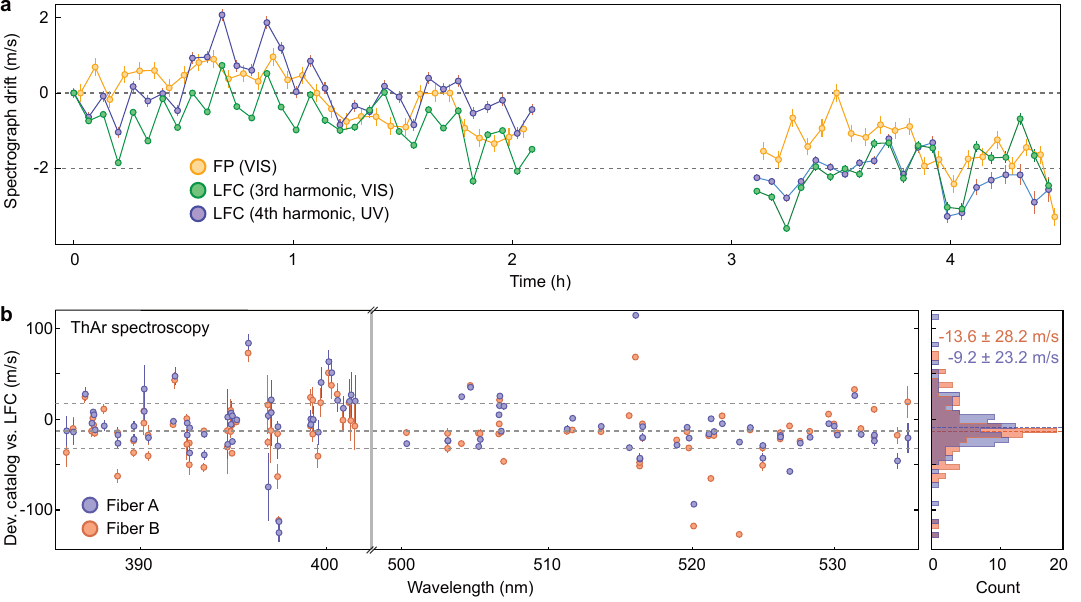}
  \caption{
    \textbf{Spectrograph calibration}.
        \textbf{a}, Spectrograph drift measurement based on the Fabry-Pérot (FP) etalon in the visible (VIS) wavelength and the laser frequency comb (LFC) in VIS and ultraviolet (UV) wavelength range. The FP calibration data has been restricted to its spectral overlap with the VIS LFC; Error bars indicate formal uncertainties based on a propagation of photon- and read-out noise.
        \textbf{b}, Absolute spectroscopy of thorium emission lines by the LFC calibrated spectrograph. The deviation of the literature values from our LFC-calibrated measurement is shown as function of  wavelength and separately for A and B fiber. Error bars indicate formal fitting uncertainties based on a propagation of photon- and read-out noise; dashed lines indicate 16th, 50th, and 84th percentiles. Right panel: histogram of the deviations, separated for Fiber~A and B. Mean and standard deviation are indicated.
 }
  \label{fig:calib}
\end{figure*}

\paragraph{Spectrograph calibration.}
With the principle of efficient harmonic generation validated, we build a dedicated ultraviolet astrocomb and use it for spectrograph calibration under realistic conditions at the high-resolution echelle spectrograph SOPHIE \citep{perruchot2008} at the Observatoire de Haute Provence (OHP, France). \rev{With the help of an Echelle-grating and a prism, the spectrograph cross disperses the input spectrum on a 2-dimensional charged coupled device (CCD) detector array so that the spectrum is arranged in a vertically separated set of nearly horizontal lines (\textit{Echelle orders}), where each line represents a distinct wavelength interval.}
The SOPHIE spectrograph, covers the wavelengths $\lambda$ from $387$~nm to $708$~nm with a resolving power of $R=\frac{\lambda}{\Delta{}\lambda}\approx75'000$ (i.e. frequency resolution of approximately 10~GHz at 400~nm wavelength) and its echelle grating is
enclosed in a pressure vessel to reduce externally induced drifts. Traditionally, wavelength calibration can be obtained via thorium-argon hollow-cathode lamps (ThAr) and a passively-stabilized Fabry-P\'erot etalon (FP) \citealt{wildi2010,wildi2012}, permitting calibration on the $1$ to $2\,\mathrm{m\,s^{-1}}$ level \citep{bouchy2013}.

As an initial IR comb generator, we utilize an 18~GHz EO comb, similar to \cite{obrzud2018} (Fig.~\ref{fig:sophie}a). A CW laser of wavelength 1556.2~nm is stabilized to an optical resonance frequency in rubidium (Rb), providing an absolute frequency anchor for the IR EO comb (Methods). Comb generation proceeds by sending the CW laser through two electro-optic phase- and one electro-optic intensity modulator, creating an initial IR comb spanning approximately 0.6~THz. The modulation frequency of 18~GHz is provided by a low-noise microwave synthesizer, referenced to a miniature Rb atomic frequency source; over the relevant time scales of our experiments, a relative frequency stability better than $10^{-11}$ is achieved (Methods). In a waveshaper, a polynomial spectral phase up to 3$^\mathrm{rd}$ order is imprinted on the EO comb spectrum, to ensure short pulses after the amplification in an erbium-doped fiber amplifier (EDFA) \rev{and an additional stretch of optical fiber for nonlinear pulse compression}. \rev{In this way, an 18~GHz femtosecond pulse train with 3~W of average power can be generated}. The pulses are out-coupled to free-space and coupled to the chip via a high-numerical aperture (NA=0.7) lens (estimated on chip pulse energy $\sim$20~pJ). Despite the high average input power levels, we did not observe any waveguide damage. 
\rev{For comb generation, we choose the waveguide in} Fig.~\ref{fig:waveguideDesign}f, \rev{as it exhibits a strong UV signal and a clear separation between the harmonics. We arrange the waveshaper and the optical fiber after the amplifier such that the driving pulses are of ca. 200~fs duration, maintaining the separation between the harmonics, so that the emergence of inconsistent frequency combs, as described above, can be excluded. As we show in the SI, Section~5, shorter pulses can enable the generation of broadband and merging harmonics, which opens opportunities for future work.
}
The generated spectra are similar to those generated at lower repetition rate (Fig.~\ref{fig:waveguideDesign}d-g).
The chip's output, \rev{including fundamental and higher-order waveguide modes}, is collimated using a parabolic mirror and sent through a tunable prism for coarse wavelength filtering and balancing \rev{of} flux levels as needed, before being coupled to a multi-mode fiber. 
\rev{The light is sent through a mode-scrambler (see Methods) and then routed to an auxiliary input port in the spectrograph's calibration unit, which marks the interface between our LFC setup and the existing telescope and spectrograph infrastructure at OHP} \citep{perruchot2008}. \rev{The calibration unit splits the comb light equally into two optical fibers (A/B fiber), which are placed next to each other on the spectrograph entrance slit, each creating an equivalent cross-dispersed spectrum on the 2-dimensional detector. }
Figure ~\ref{fig:sophie}b shows a comb spectrum as recorded on the CCD detector of the \rev{SOPHIE spectrograph}. \rev{The comb structure of the LFC spectrum is clearly visible along the individual spectral orders of the cross-dispersed echelle spectrograph for both A and B fibers.} 
Data \rev{extraction} and reduction of the LFC exposures is performed using a custom code initially developed for the {ESPRESSO} spectrograph \citep{schmidt2021}. It \rev{extracts the spectral information from the 2-dimensional detector frame} using a variant of the \textit{flat-relative optimal extraction} algorithm \citep{zechmeister2014}, provides default wavelength \rev{calibration} based on a combined ThAr/FP wavelength solution \citet{cersullo2019,schmidt2021}, \rev{and combines the data from the individual spectral orders to a single} 1-dimensional spectrum (Fig.~\ref{fig:sophie}c) \rev{and thus makes it accessible to subsequent analysis}.
The resolving power of the spectrograph is not sufficient to fully separate the comb lines. In particular in the UV part of the spectrum, the wings of the spectrograph's line-spread functions for adjacent comb lines  overlap and imply a reduced contrast. Nevertheless, the individual comb lines are clearly discernible. 

To test the \textit{stability} of the LFC-based wavelength calibration, we obtain over several hours repeated LFC exposures interleaved with exposures of the FP etalon. The two types of exposures allow to measure the spectrograph drift in independent ways, once relative to the LFC and once relative to the FP.
After data reduction as described above, all LFC lines in the spectra are fitted individually to determine their positions on the detector (Methods).
For the FP exposures, the gradient method \citet{bouchy2001} is used to non-parametrically detect shifts of the spectra on the detector.
As shown in Figure~\ref{fig:calib}a, we obtain separate drift measurements from the 3$^\mathrm{rd}$ and 4$^\mathrm{th}$ harmonic and in addition one drift value from the FP, where we only use the FP lines that overlap spectrally with the 3$^\mathrm{rd}$ harmonic (the FP calibration does not cover the UV). 
All three measurements agree well \rev{(within the amplitude of their scatter)} and detect a \rev{systematic} spectrograph drift of ca. \rev{-2~m/s} over 4~hours. 
The agreement between the 3$^\mathrm{rd}$ harmonic and the FP validates the LFC, and the agreement between the 4$^\mathrm{th}$ harmonic with the two other measurements extends the validation to the UV regime.
The slightly higher noise in the LFC data, \rev{i.e. remaining differences between} 3$^\mathrm{rd}$ and 4$^\mathrm{th}$ \rev{harmonic and scatter around the global trend, compared to the FP drift measurement,} can probably be attributed to a combination of effects, e.g. intrinsic systematics of the spectrograph like detector stitching issues \cite{coffinet2019} or charge transfer inefficiency (CTI) \cite{bouchy2009} combined with a change in the flux levels of the LFC lines. \rev{Moreover, the line spacing of our comb (18~GHz), which is limited by our specific technical components, is less than half the FSR of the FP calibrator (39~GHz). As a result, comb line fitting (cf. Methods) is more challenging and slightly higher scatter can be expected, when compared to the FP-based calibration. Implementing a comb with larger spacing or higher resolution spectrographs can address this challenge. The fact that nevertheless the scatter in the FP and comb-based calibrations are comparable, indicates that the obtained precision is here limited by the spectrograph, not by the comb (indeed, the scatter is consistent with the design goals of the spectrograph} \cite{perruchot2008}). 

\rev{To test the \textit{accuracy} of our LFC-based calibration, we compare it to the  established laboratory wavelengths of the ThAr hollow-cathode lamp. For this, two exposures, one containing the LFC and the other the ThAr spectrum, are obtained in quick succession, which ensures that no relevant spectrograph drift happens between the two exposures.}
\rev{In the extracted LFC spectrum, individual lines are fitted (see above and Methods) and, together with the knowledge about their true, intrinsic frequencies }(Eq.~\ref{eq:comb_freqs}),\rev{ a purely LFC-based wavelength solution, i.e. a relation between CCD pixel position and wavelength, is derived for the spectral region covered by the comb. 
This LFC wavelength solution is then applied to the extracted ThAr spectrum and, after fitting of line centroids, LFC-calibrated wavelentghs are assigned to the thorium lines. These are then compared to the laboratory wavelengths from }\cite{redman2014} \rev{and the deviations are shown in }Fig.~\ref{fig:calib}b.
\rev{The histogram of all deviations shows a global offset of $-9$ to $-14$~m/s with a standard deviation of $>20$~m/s, which is consistent with zero offset, and also} with an equivalent comparison performed with the highly accurate ESPRESSO spectrograph \cite{schmidt2021}. \rev{We note that the large uncertainty (the standard deviation in the histogram) is not limited by the comb calibrator but due to the low number of only $\sim$100 ThAr lines that are available for comparison.
The scatter of individual line deviations} (and the difference between Fibers A and B), which significantly exceeds the formal uncertainties, has also been observed for other spectrographs (e.g. ESPRESSO \cite{schmidt2021}) and seems to be a common systematic, probably related to an imperfect determination of the instrumental line-spread function.
We confirm that this scatter is identical when comparing thorium lines to LFC or ThAr/FP wavelength solutions \citet{cersullo2019,schmidt2021}.
Therefore, we conclude that 
the LFC wavelength calibration is accurate to the level testable with SOPHIE.
Both measurements combined validate on the proof-of-concept level that the generated astrocomb can provide precise and absolute calibration even at UV wavelengths.

\paragraph{Microresonator based UV astrocombs.}
\begin{figure}[t!]
  \centering
  \includegraphics[width=\columnwidth]{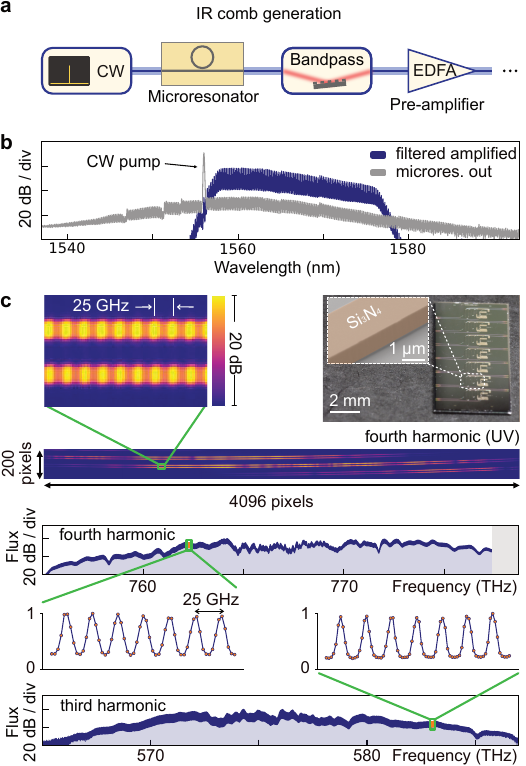}
  \caption{
    \textbf{Microresonator-based ultraviolet astrocombs}.
        \textbf{a}, Modified setup for microresonator-based ultraviolet (UV) astrocomb generation.
        \textbf{b}, Microresonator comb output spectrum (gray); bandpass filtered and pre-amplified microresonator-based comb spectrum.
        \textbf{c}, Microresonator-based UV fourth harmonic and visible (VIS) third harmonic comb spectra, following the scheme in Fig.~\ref{fig:sophie}.
        The photo shows the photonic chip with the microresoantor (dashed-line box) and the inset shows a scanning electron microscope image of the Si$_3$N$_4$ resonator waveguide. 
 }
  \label{fig:microres}
\end{figure}
Finally, we explore photonic chip-integrated microresonators (\textit{microcombs} \cite{kippenberg2018, pasquazi2018, gaeta2019}) as the fundamental IR comb generator. 
While their current level of maturity is lower than that of EO comb generators or mode-locked lasers, microcombs can provide compact and energy efficient \cite{helgason2022} high-repetition rate, low-noise femtosecond pulses \cite{herr2014}, and their potential for microresonator astrocombs has already been demonstrated in the IR spectral range \cite{suh2019,obrzud2019b}.
Combined with highly efficient frequency conversion in LiNbO$_3$ waveguides, potentially in a resonant configuration \cite{herr2018, jankowski2021, roy2022}, they could open new opportunities for space-based calibrators, where energy consumption, size, and robust integration are key. Such space-based systems could calibrate ground-based observatories through the atmosphere, or provide calibration to space-based observatories, which would entirely avoid limiting atmospheric effects, such as absorption or turbulence.
Therefore, as a proof-of-concept, we replace the electro-optic comb generator with a silicon nitride microresonator \cite{ye2019, ye2022} (Fig.~\ref{fig:microres}c, photo) operating at  $f_\mathrm{rep}$=25~GHz (Methods). A portion of the generated comb spectrum is bandpass-filtered and pre-amplified in an erbium-doped fiber amplifier (cf. Fig.~\ref{fig:microres}a). Figure~\ref{fig:microres}b shows the IR comb spectrum generated by the microresonator as well as the filtered and amplified spectrum. The remainder of the setup is the same as described before for the EO astrocomb (Fig.~\ref{fig:sophie}a).
While we do not stabilize the microcomb spectra 
we observe broadband harmonics of distinct comb lines, similar to the EO comb configuration (Fig.~\ref{fig:microres}c), illustrating the potential of implementing all nonlinear optical components via highly-efficient photonic integrated circuits.

\rev{Although lithium niobate is known to show photo-induced damage by visible and ultraviolet light, we did not observe any signs of degradation in our experiments and attribute this to the relatively low flux levels in these wavelength domains (The power in the harmonics and corresponding conversion efficiencies are presented in the SI, Section~3). In a dedicated experiment, described in Section~4 of the SI, we transmitted light of a 405~nm laser diode through the waveguide and did also not observe any degradation. Zirconium-doping of the lithium niobate can further increase the damage threshold for visible and ultraviolet light} \cite{liu2010}.


\subsection*{Discussion}
Our results show for the first time astronomical spectrograph calibration in the UV wavelength range with a laser frequency comb. This demonstration is particularly relevant to the emerging field of optical precision cosmology with next generation instruments and telescopes. Our astrocomb's architecture is based on periodically-poled lithium niobate nanophotonic waveguides tailored to provide efficient frequency conversion of a robust, alignment-free 18~GHz electro-optic frequency comb. Moreover, we show compatibility of the approach with photonic chip-integrated microcombs, creating opportunities for future space-based systems. This also represents the first demonstration of nonlinear transfer of a microcomb to VIS and UV wavelengths, adding new operating domains to the repertoire of these sources. \rev{The approach demonstrated here can be utilized to extend calibration beyond established wavelength ranges towards shorter wavelengths.}
\rev{Future work may pursue wider harmonic spectra (cf. SI, Section~5), where however, consistency of the generated spectra via suitable control of the fundamental comb's offset-frequency (as described above) and noise suppression as previously demonstrated for electro-optic combs \cite{beha2017} would need to be implemented. With improved chip-coupling \cite{he2019, churaev2022} this could enable gap-free VIS and UV operation based on robust telecommunication wavelength IR lasers.}
Similar results may be possible also in other waveguide materials such as aluminum nitride (AlN), where broadband phase matching may be achieved through chirped geometries \cite{liu2019}. \rev{In addition to the development of the comb source itself, spectral flattening \cite{probst2014} will be needed to ensure optimal exposure of the CCD and to avoid systematic effects that can arise from strong modulation of the combs spectral envelope.}
In sum, the demonstrated approach opens a viable route towards astronomical precision spectroscopy in the ultraviolet based on robust infrared lasers and may eventually contribute to major scientific discoveries in cosmology and astronomy.


\subsection*{Methods}
\small
\paragraph{Waveguide and microresonator fabrication.}
Lithium niobate waveguides for UV comb generation are fabricated from a 800~nm-thick x-cut LiNbO\textsubscript{3} layer on 3~µm SiO\textsubscript{2} and bulk Si substrate (NANOLN). First, chromium \rev{(Cr)} electrodes for periodic poling are \rev{fabricated} using electron beam lithography and lift-off. Ferroelectric domain inversion is performed by applying \rev{a high-voltage waveform across the electrodes, adapted from} \cite{younesi_periodic_2021} \rev{(SI,  Fig.~5)}. After poling, the electrodes are removed via chemical etching. Waveguides are patterned using electron beam lithography on a \rev{Cr} hard mask. Cr is etched using \rev{argon (Ar)} ion-beam etching \rev{and} subsequently, the lithium niobate layer is fully etched (no slab remaining) via reactive-ion etching with fluorine chemistry. The remaining Cr hard mask is later removed \rev{with chemical etching.} \rev{Lastly}, a 3~µm-thick SiO\textsubscript{2} \rev{cladding} layer is deposited using chemical vapor deposition \rev{and waveguide} facets are defined by deep etching to ensure low-loss coupling to the waveguides. \rev{More details regarding the fabrication of lithium niobate waveguides can be found in SI, Section~2}. The Si\textsubscript{3}N\textsubscript{4} microresonator is designed with a finger shape \cite{ye2022} and fabricated via the subtractive processing method as described in ref.~\cite{ye2019}. The height and the width of the SiN waveguide are 740~nm and 1800~nm respectively, which result in a group velocity dispersion coefficient of $\beta_2$ = --73~ps$^2$/km for the fundamental TE mode at the pump wavelength. The total linewidth of the critically coupled resonator is 25~MHz. Lensed fibers are used for coupling into and out of the bus waveguide. More details are provided in Section~2 of the SI.

\paragraph{CW laser frequency lock.}
A narrow-linewidth CW laser at telecommunication wavelength 1556.2~nm (Redfern Integrated Optics, RIO Planex) is split and one part is used for EO comb generation. The other part is frequency-doubled and brought into resonance with the $5S_{1/2} (F_g=3) - 5D_{5/2} (F_e=5)$ two-photon transition in $^{85}$Rb atoms in a microfabricated atomic vapor cell, similar to \cite{newman2021}. 
A stable frequency lock of the CW laser to the two-photon transition is achieved by means of synchronous detection and by modulating the laser’s injection current. Using the same vapor cell and a laser with higher phase noise, we measured the fractional frequency instability of our system to an Allan deviation better than $2 \times 10^{-12}$ for averaging times from 1 s to $10^4$ s. 

\paragraph{Miniature atomic frequency source.}
The 10 MHz sine wave output of a commercial miniature atomic frequency source is provided to the microwave synthesizer as a stable reference. The miniature atomic clock we use is based on a transition in the ground state of $^{85}$Rb. Its relative frequency stability is specified by the manufacturer to be better than $1 \times 10^{-10}$ ($1 \times 10^{-11}$) at 1~s (100~s) averaging time; daily drift $< 1 \times 10^{-11}$.

\paragraph{Mode scrambler and calibration unit.}
The mode-scrambler suppresses detrimental laser speckles on the CCD (modal noise) that would result from the coherent LFC illumination after propagating through \rev{the multi-mode fiber train of the SOPHIE spectrograph}. It consists of a rotating diffuser disc and has a throughput of $\approx5\,\%$ effectively washing out any laser speckle pattern. Averaged over one CCD exposure  time, it populates all fiber modes and \rev{therefore ensures} a speckle-free image of the \rev{spectrographs's} entrance aperture on the CCD. After the mode scrambler the light is guided by a multi-mode fiber to the calibration unit. The calibration unit selects one of the installed calibration light sources or, via a temporarily added additional input port, the light from the LFC. The selected light is then sent to the spectrograph's front-end, \rev{mounted at the Cassegrain focus of the OHP 1.93\,m telescope}, where it is \rev{simultaneously} injected into A and B fibers feeding the spectrograph \citep{perruchot2008}. 

\paragraph{LFC line fitting and computation of drift.} 
The LFC lines are fitted to find their exact center positions. Due to the limited resolution of the spectrograph, the comb lines are not fully separated \rev{but partially overlap, particularly in the UV domain}. \rev{Therefore, always three LFC lines are fitted together}, a central line of interest and two flanking ones to model the flux contributed by their wings. \rev{In addition, the diffuse background flux is locally modeled by a polynomial of first order. A detailed characterization of the SOPHIE instrumental line-spread function is not available. Therefore, we have to resort to model all lines with a simple Gaussian shape.}
\rev{In addition, due to the limited sampling (about 2.5\,pixel per FWHM) on the detector, a simultaneous and unconstrained fit of all nine model parameters (positions and amplitudes of the thee lines in question, one common line width, and amplitude and slope for the diffuse background) leads to degeneracies and unphysical fit results.
The width of all LFC lines in the fit is therefore fixed to a FWHM of 2.47~pixels. This value was inferred from the thorium lines and, although not capturing the variation of the instrumental line-spread function along individual spectral orders or with wavelength in general, found to better describe the line shapes than a width that is fixed in velocity units.}

For each exposure, the fitted line centroids are compared to the corresponding positions in the first exposure of the sequence, the differences converted to a radial velocity shifts following the Doppler formula, and then combined to provide one average drift measurement per exposure.



\subsection*{Funding}
\small
This project has received funding from the Swiss National Science Foundation (Sinergia BLUVES CRSII5\_193689), the European Research Council (ERC) under the EU’s Horizon 2020 research and innovation program (ERC StG 853564 and ERC CoG 771410), and through the Helmholtz Young Investigators Group VH-NG-1404; the work was supported through the Maxwell computational resources operated at DESY. The lithium niobate waveguides were fabricated in the EPFL Center of MicroNanoTechnology (CMi).


\subsection*{Data availability}
\small
The datasets generated and analysed during the current study are available from the corresponding author on reasonable request.

\subsection*{Code availability}
\small
The numeric simulations in the current study use an open source code available at https://github.com/pychi-code/pychi.


\subsection*{Competing interests}
\small
We declare that none of the authors have competing interests.




\printbibliography
\newpage

\onecolumn
\normalsize

\renewcommand{\figurename}{Supplementary Figure}

\title{Ultraviolet astronomical spectrograph calibration with \\laser frequency combs from nanophotonic lithium niobate waveguides
        \\ \vspace{1cm} Supplementary Information}

\author{Markus Ludwig,\authormark{1,*}
        Furkan Ayhan,\authormark{2,*}
        Tobias M. Schmidt,\authormark{3,*} \\
        Thibault Wildi,\authormark{1} 
        Thibault Voumard,\authormark{1} 
        Roman Blum,\authormark{4}
        Zhichao Ye,\authormark{5}   
        Fuchuan Lei,\authormark{5} \\
        François Wildi,\authormark{3} 
        Francesco Pepe,\authormark{3} 
        Mahmoud A. Gaafar,\authormark{1}  
        Ewelina Obrzud,\authormark{4} 
        Davide Grassani,\authormark{4} \\
        Olivia Hefti,\authormark{4} 
        Sylvain Karlen,\authormark{4} 
        Steve Lecomte,\authormark{4}
        François Moreau,\authormark{6}
        Bruno Chazelas,\authormark{3}
        Rico Sottile,\authormark{6} \\
        Victor Torres-Company,\authormark{5}
        Victor Brasch,\authormark{7} 
        Luis G. Villanueva,\authormark{2}
        François Bouchy,\authormark{3}
        Tobias Herr\authormark{1,8,**}
}

\date{%
    \small $^1$Deutsches Elektronen-Synchrotron DESY, Notkestr. 85, 22607 Hamburg, Germany \\
    \small $^2$École Polytechnique Fédérale de Lausanne (EPFL), 1015 Lausanne, Switzerland \\
    \small $^3$ Observatoire de Gen\`eve, Département d’Astronomie, Universit\'e de Gen\`eve, Chemin Pegasi 51b, 1290 Versoix, Switzerland \\
    \small $^4$ Swiss Center for Electronics and Microtechnology (CSEM), 2000 Neuch\^atel, Switzerland\\
    \small $^5$ Department of Microtechnology and Nanoscience, Chalmers University of Technology, 41296 Gothenburg, Sweden \\
    \small $^6$ Observatoire de Haute-Provence, CNRS, Université d'Aix-Marseille, 04870 Saint-Michel-l'Observatoire, France \\
    \small $^7$ Q.ANT GmbH, Handwerkstraße 29, 70565 Stuttgart, Germany\\
    \small $^8$Physics Department, Universität Hamburg UHH, Luruper Chaussee 149, 22607 Hamburg, Germany\\
    \small $^*$These authors contributed equally.\\
    \small $^{**}$tobias.herr@desy.de
}

\maketitle

\setcounter{figure}{0}
\setcounter{page}{1}
\begin{refsection}

\section{Linewidth and background in the harmonic frequency comb}

The lineshape of a comb line is defined by its phase noise spectrum. Phase noise may broaden the linewidth or may contribute to a broadband background signal. Both linewidth and background have important implications for astrocombs and are discussed below.

\subsection{Phase noise and comb lineshape}
\label{SI:sec:phase_noise}
Both $\nu_0$ and $f_\mathrm{rep}$, which are the defining frequencies of the comb (Eq.~1, main text), exhibit non-zero phase noise described by their (single-sided) phase noise power spectral densities (PSDs) $S_{\phi}^{\nu_0}(f)$ and $S_{\phi}^{f_\mathrm{rep}}(f)$, where $f$ is the offset frequency from the respective carrier wave. The resulting phase noise of the comb lines is 
\begin{equation}
    S_{\phi}^{\nu_{m,n}} = m^2S_{\phi}^{\nu_0} + n^2S_{\phi}^{f_\mathrm{rep}}\, ,   
\end{equation}
where $m=1,2,..$ indicates the harmonic and $n=\pm1,\pm2,..$ the line index relative to the center of the harmonic. 
The second term usually limits the useful span (i.e., a maximal $|n|$) of an electro-optic comb generator. To increase the useful span, noise-filtering can be applied to suppress the emergence of a strong background signal as well as linewidth broadening \cite{beha2017}. 

Creating broadband comb spectra via cascaded $\chi^{(2)}$-harmonics, rather than $\chi^{(3)}$-based broadening around $\nu_0$, shows distinct noise properties as frequencies $\nu$ much higher than $\nu_0$ can be reached through lines in a harmonic $m$ while keeping $n << (\nu_{m,n}-\nu_0)/f_\mathrm{rep}$. Neglecting the small impact of $m^2 S_{\phi\phi}^{\nu_0}$ (as $m<<n$ ), the phase noise of comb lines of frequency $\nu_{m,n}$ that are part of the m$^{th}$ harmonic comb does not grow in function of their frequency difference from $\nu_0$, but only with their distance from the m$^{th}$ harmonics center frequency $\nu_{m,0}$. This permits generating UV astrocombs in a limited span around the fourth harmonic center frequencies $\nu_{4,0}$ from a near-infrared electro-optic comb generator without noise-filtering;  wider spectra around the harmonics will still require noise filtering. The different noise properties of $\chi^{(2)}$ and $\chi^{(3)}$-based spectra are illustrated in Supplementary Fig.~\ref{fig:phase_noise_schematic}.

\begin{figure}[t!]
  \centering
    \includegraphics[width=0.5\columnwidth]{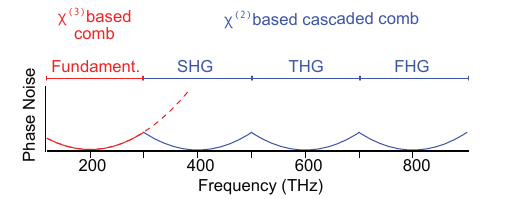}
  \caption{
    \textbf{Phase noise} in a traditional $\chi^{(3)}$-based comb compared to the phase noise a $\chi^{(2)}$-based comb generated through cascaded harmonics..
 }
 \label{fig:phase_noise_schematic}
\end{figure}

Based on the phase noise $S_\phi^{f_\mathrm{rep}}$ we can (numerically) estimate the lineshape $S_n(\nu_\mathrm{rel})$ of a comb line that is $n$ lines away from the center of the harmonic by computing the Fourier transform ($\nu_\mathrm{rel}$ is the optical frequency relative to the center of the optical line)
\begin{equation}
    S_n(\nu_\mathrm{rel}) = 2 \int_{-\infty}^{\infty} \Gamma(\tau) \exp\left[-i2\pi\nu_\mathrm{rel}\tau\right] \, \mathrm{d}\tau
\end{equation}
of the normalized autocorrelation function \cite{domenico2010SimpleApproachRelation}
\begin{equation}
    \Gamma(\tau) = \exp\left[-2\int_0^{\infty}n^2S_\phi^{f_\mathrm{rep}}(f) \sin^2(\pi f\tau)\, \mathrm{d}f\right] \, .
\end{equation} 

\subsection{Experimental observation of line contrast and comparison with numeric model}

\begin{figure}[ht!]
  \centering
    \includegraphics[width=0.9\columnwidth]{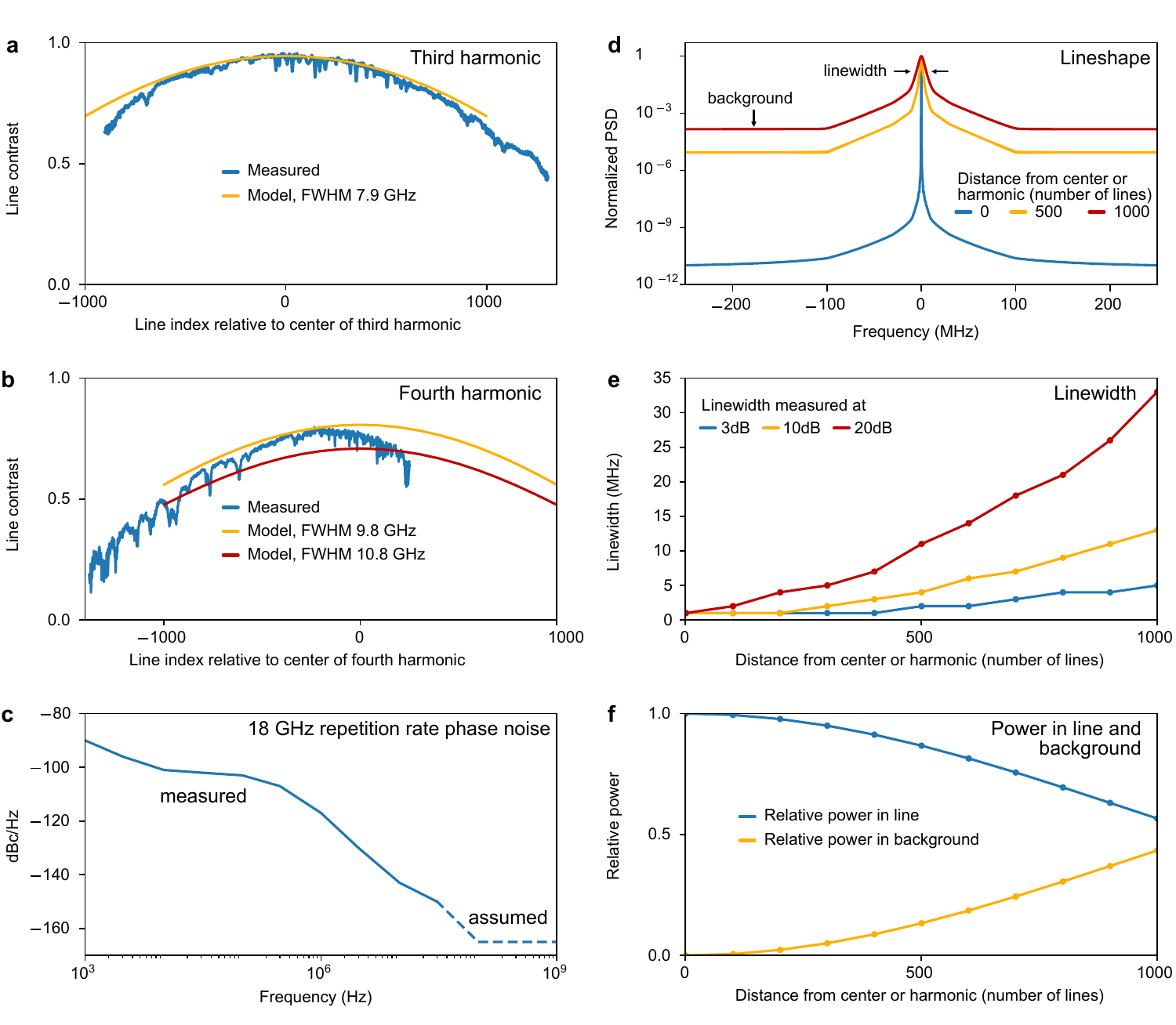}
  \caption{
    \textbf{Contrast and lineshape. a, } Measured line contrast of the third harmonic and predicted contrast based on the model described in the text for an instrumental line-spread function with a full-width-at-half-maximum (FWHM) of 7.9~GHz. \textbf{b, } Same as in panel a, however, for the fourth harmonic and assuming an instrumental line-spread function with FWHM of 9.8~GHz and 10.8~GHz. \textbf{c, } Power spectral density $S_\phi ^{f_\mathrm{rep}}$ of the repetition rate phase noise for our model, based on the measured phase noise of the microwave synthesizer driving the electro-optic comb up to 30~MHz frequency and assumed phase noise above 30~MHz (see text for details). \textbf{d, } Comb lineshape (normalized PSD) for the phase noise in panel c for different distances from the center of a harmonic. \textbf{e,} 3dB, 10dB and 20dB linewidth of the comb lines in dependence of their distance from the center of a harmonic.  \textbf{f,} Relative power contained in the comb line and in the flat background (see text for details) in dependence of their distance from the center of a harmonic.
 }
  \label{fig:exp_contrast_lineshape}
\end{figure}

An astronomical spectrograph cannot resolve the lineshapes of the comb lines but instead observes the convolution of the comb spectrum with the spectrograph's instrumental line-spread function (LSF). Depending on the lineshape of the comb lines (linewidth and background), the spacing of the comb lines and the LSF, comb lines may be observed by the spectrograph with a reduced level of contrast. Based on the discussion in Supplementary Section~\ref{SI:sec:phase_noise} one would expect that the observed line contrast reduces 
for larger $|n|$ due to the increase of phase noise (causing increased background and linewidth). This is indeed observed in the experimental data. Supplementary Figure~\ref{fig:exp_contrast_lineshape}a,b, shows the line contrast (extracted from main text Figure~3 as the peak-valley difference divided by the peak height) in dependence of the line index $n$ relative to the spectral centers of the third and fourth harmonics. As expected, the contrast is highest in the center of the harmonics and decreases with spectral distance from the harmonic center.
Under the assumption that in the center of the harmonics the lines are narrow and there is negligible background, we can estimate the full-width-at-half maximum (FWHM) of the instrumental line-spread function which we assume to be of Gaussian shape. We find $\approx7.9$~GHz for the FWHM in the spectral region of the third harmonic and 9.8~GHz to 10.8~GHz in the spectral region of the fourth harmonic. Assuming that the PSD of the repetition rate phase noise can be approximated at noise frequencies up to 30~MHz by the phase noise of our microwave synthesizer and above 30~MHz exhibits a white phase noise floor of $-165$~dBc/Hz (see ~\ref{fig:exp_contrast_lineshape}c ), we derive the lineshapes according to Supplementary Section~\ref{SI:sec:phase_noise} and perform the convolution with the estimated instrumental line-spread functions. These model results are overlayed in the Supplementary Figure~\ref{fig:exp_contrast_lineshape}a,b and show good qualitative agreement with the observed data. For clarity we note that the experimental noise floor in $S_\phi^{f_\mathrm{rep}}$ may not only stem from the microwave source but also emerge from amplified spontaneous emission (ASE) in the erbium-doped fiber amplifiers or other noise sources.
To better illustrate the underlying comb lineshapes (prior to convolution), Supplementary Figure~\ref{fig:exp_contrast_lineshape}d shows their characteristic shapes for different values of $n$. Panels e and f show the linewidth at different levels as well as the relative power contained in the narrow line and in the flat background as function of $n$. The background power is determined by multiplying the background level in the optical PSD with the comb line spacing $f_\mathrm{rep}$. Subtracting the background power from the total power yields the line power.

\subsection{Comparison of linewidth and line/background power in $\chi^{(2)}$- and $\chi^{(3)}$-based combs}

\begin{figure}[ht!]
  \centering
    \includegraphics[width=0.9\columnwidth]{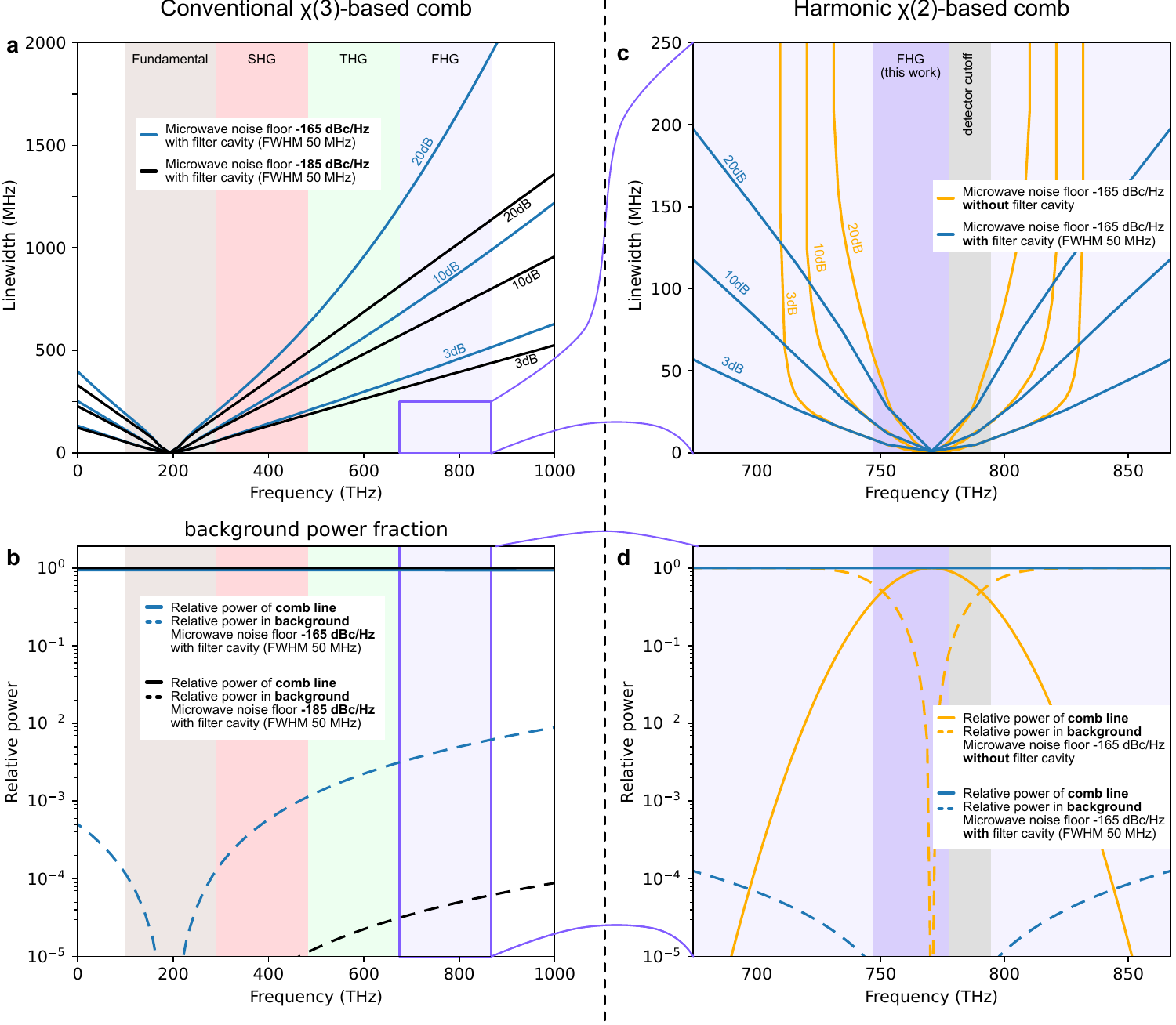}
  \caption{
    \textbf{Comparison of key characteristics in $\chi^{(2)}$- and $\chi^{(3)}$-based electro-optic combs, a, } 3dB, 10dB and 20dB linewidth for a $\chi^{(3)}$-based electro-optic comb with cavity noise supression (Full-width-at-half-maximum, FWHM, 50~MHz). The microwave phase noise is assumed to be as in Supplementary Figure~\ref{fig:exp_contrast_lineshape} with a noise floor of $-165$~dBc/Hz and $-185$~dBc/Hz, respectively. The color-coded background indicates the spectral domains of fundamental comb, second, third and fourth harmonic generation (SHG, THG, FHG) \textbf{b, } Relative power in the flat background and in the comb line for the conditions as in panel a. \textbf{c, } 3dB, 10dB and 20dB linewidth for a $\chi^{(2)}$-based electro-optic comb around the center of the fourth harmonic with and without a noise filter cavity and a microwave phase noise floor of $-165$~dBc/Hz. The spectral range used in this work and the detector cutoff are indicated. \textbf{d, } Relative power in the flat background and in the comb line for the conditions as in panel c.
 }
\label{fig:comparison_chi2_chi3}
\end{figure}

To provide additional information, also with regard to future implementations of UV electro-optic astrocombs, we have added a detailed comparison of $\chi^{(2)}$- and $\chi^{(3)}$-based comb generation in Supplementary Figure~\ref{fig:comparison_chi2_chi3}. This figure shows in the left column (panels a,b) the conventional $\chi^{(3)}$-based approach and in the right column (panels c,d) the approach based on $\chi^{(2)}$ processes. In both cases we assume the repetition rate phase noise as shown in Figure~\ref{fig:exp_contrast_lineshape}c, where for the $\chi^{(3)}$-based approach we also show a scenario with a noise floor at $-185$~dBc/Hz (such low noise floor could be achieved when the synthesizer outputs at least 11~dBm of microwave power (thermal noise floor at $-174$~dBm/Hz); higher power is needed when additional noise, e.g. from microwave amplification, is relevant).
As a comparison between panels a and c shows, the width of the comb lines generated in the $\chi^{(2)}$-based approach can be well below those generated in a $\chi^{(3)}$-based approach. Over the wavelength range utilized in our work, and with the parameters assumed here, the $\chi^{(2)}$-generated lines, even without noise suppression, are expected to be narrower than those one would expect from a $\chi^{(3)}$-based approach with a noise suppression filter (Here we assume a noise filter of 50~MHz width). Figure~\ref{fig:comparison_chi2_chi3} panels b and c compare the relative power that is contained in the line and in the flat background.
The spectral interval utilized in our work is highlighted in panels c and d, indicating a sufficiently low contribution of the background so that the comb lines can be observed. Noise filtering can provide effective means of suppressing the background flux both for the  $\chi^{(3)}$-based approach as well as for extending the  $\chi^{(2)}$-based approach to wider bandwidth.

\section{Fabrication of periodically-poled lithium niobate waveguides}

\begin{figure}[ht!]
  \centering
  \includegraphics[width=80mm]{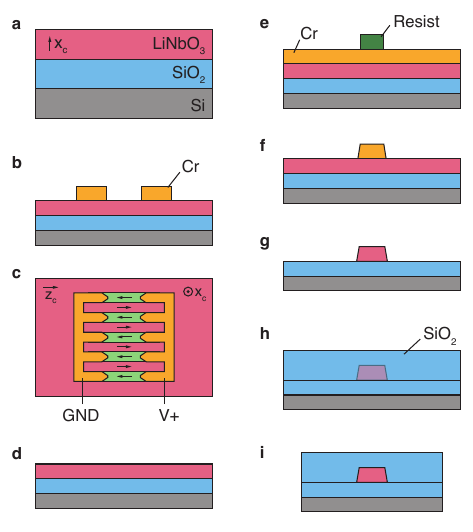}
  \caption{
    \textbf{Fabrication of periodically-poled lithium niobate wavegeuides.}
        \textbf{a,} LNOI substrate with 800 nm x-cut LiNbO\textsubscript{3}, 3 µm SiO\textsubscript{2} and bulk Si.
        \textbf{b,} Deposition of Cr electrodes for periodic poling with electron-beam lithography, electron-beam evaporation and lift-off.
        \textbf{c,} Periodic poling with a high-voltage source. Sample is illustrated from top.
        \textbf{d,} Removal of the Cr electrodes after periodic poling.
        \textbf{e,} Cr hard mask deposition with electron-beam evaporation and patterning of the waveguides with electron-beam lithography.
        \textbf{f,} Ar ion-beam etching of the Cr hard mask.
        \textbf{g,} Fluorine reactive-ion etching of the LiNbO\textsubscript{3}.
        \textbf{h,} SiO\textsubscript{2} cladding deposition with chemical vapor deposition.
        \textbf{i,} Series of reactive ion etching steps to etch SiO\textsubscript{2}, LiNbO\textsubscript{3} and Si layers to reveal waveguide facets.}
  \label{fig:fabrication}
\end{figure}

Lithium niobate waveguides for UV comb generation are fabricated on 16-by-16 mm$^2$ chips diced from a 100 mm wafer (NANOLN) with 800~nm-thick x-cut LiNbO\textsubscript{3} layer on 3~µm SiO\textsubscript{2} and bulk Si substrate (Supplementary Figure \ref{fig:fabrication}a). To ensure precise alignment of the waveguides and poling electrodes in subsequent lithography steps to the crystal axes of the lithium niobate, platinum (Pt) alignment marks are patterned on wafer scale before dicing. This is done with electron-beam lithography (EBL), electron-beam evaporation (EBE) and a lift-off process.

\begin{figure}[ht!]
  \centering
  \includegraphics[width=120mm]{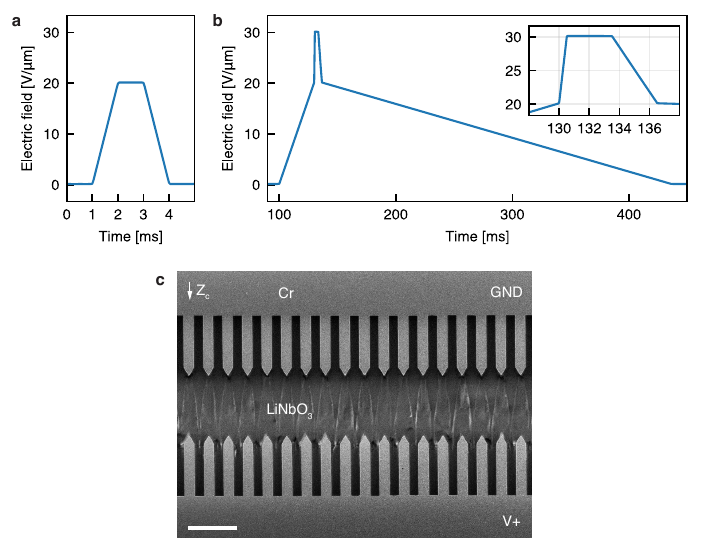}
  \caption{
    \textbf{Periodic poling of the lithium niobate waveguides}.
        \textbf{a,} One of three high-voltage \textit{pre-poling} signals applied to electrodes for periodic poling. 
        \textbf{b,} High-voltage \textit{main} signal applied for poling. Inset shows the detailed section of the peak poling signal.
        \textbf{c,} SEM image of the poled lithium niobate sample. Scale bar is 10 µm.}
  \label{fig:poling}
\end{figure}

First, chromium (Cr) poling electrodes are patterned via EBL, EBE and lift-off using MMA/PMMA bilayer resist (Supplementary Figure \ref{fig:fabrication}b). Electrodes are patterned such that the applied electric field across the electrodes is along the z-axis of lithium niobate (Supplementary Figure \ref{fig:fabrication}c). Ferroelectric domain inversion is performed by applying a high-voltage (HV) field accross the electrodes in two stages: First, a \textit{pre-poling} signal is repeated three times to help nucleation of domains closer to positive (V+) electrode (Supplementary Figure \ref{fig:poling}a). Secondly, a higher field (30 V/µm), followed by a slow decay is applied to propagate the iniated domains to the ground (GND) electrode (Supplementary Figure \ref{fig:poling}b). After poling, samples are inspected under a scannning electron microscope (SEM) (Supplementary Figure \ref{fig:poling}c) before removing the electrodes with wet etching (Supplementary Figure \ref{fig:fabrication}d).
Before waveguide fabrication, a Cr layer is deposited using EBE to be used as a hard mask for etching lithium niobate. Later, waveguides are patterned with EBL (Supplementary Figure \ref{fig:fabrication}e). The Cr hard mask is etched using argon (Ar) ion-beam etching (IBE) (Supplementary Figure \ref{fig:fabrication}f). Subsequently, the lithium niobate layer is fully etched with reactive-ion etching (RIE) with fluorine chemistry (CHF\textsubscript{3}/Ar) (Supplementary Figure \ref{fig:fabrication}g). After etching, the remaining Cr hard mask is removed and the structures are cleaned with wet etching. 
Lastly, samples are cladded with a 3~µm-thick SiO\textsubscript{2} layer, deposited via chemical vapor deposition (CVD) (Supplementary Figure \ref{fig:fabrication}h). Finally, waveguide facets are defined by deep etching of SiO\textsubscript{2}, LiNbO\textsubscript{3} and Si layers to ensure better coupling to the waveguides.

\pagebreak

\section{Power consumption and nonlinear conversion efficiency}
\subsection{Power consumption}
The energy consumption of the electro-optic comb generator is dominated by the multi-W erbium-doped fiber amplifier (EDFA) and the microwave electronics. Assuming a 10\% wall-plug efficiency the EDFA will approximately consume 35~W of electrical power. The microwave synthesizer consumes under maximal load, depending on the specific hardware, ca. 80-130~W of electrical power, the microwave amplifiers jointly ca. 50~W of electrical power. In sum this amounts to approximately 200~W of electrical power consumption, about the same as a standard desktop computer. In case of the microresonator-based approach, one can subtract the power consumption of the microwave source and microwave amplifier, resulting in a power consumption of approximately 35~W. 
The power consumption of our mode-locked laser is below 15~W. However, considering the $\sim$200-times lower repetition rate, the energy consumption per pulse is not better than of the electro-optic comb, although it reaches higher per pulse energy.

\subsection{Nonlinear conversion efficiency}
To derive the conversion efficiency of the fundamental comb into the harmonics, the experimental spectra shown in Figure~2g in the main text complemented by additional spectra at different pulse energies were integrated over the respective spectral ranges of the harmonic orders. The left panel of Supplementary Figure~\ref{fig:ConversionEfficiency} shows the resulting output power levels of the fundamental and the harmonics up to the fourth order as a function of on-chip input pulse energy. The log-log scale clearly reveals that the slopes increase with harmonic order. Consistently, also the pulse energy threshold to cross the noise floor of the spectrum analyzer increases with the harmonic order. The fact that the power in the fundamental saturates beyond 40~pJ and eventually even drops can be attributed both to an increasing depletion into the harmonics and additional spectral broadening of the fundamental which eventually exceeds the range detected by the spectrum analyzer (1750~nm). 
The right panel of Supplementary Figure~\ref{fig:ConversionEfficiency} shows the conversion efficiency, i.e., the ratio of the power contained in the respective harmonic orders vs. the power in the fundamental comb. A conversion efficiency well exceeding 0.1~\% is achieved for the fourth harmonic. 
\begin{figure}[ht!]
  \centering
  \includegraphics[width=\textwidth]{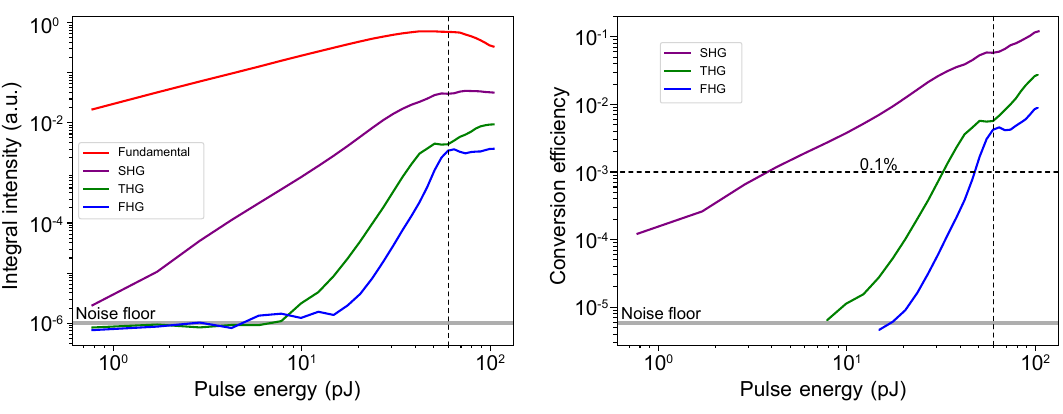}
  \caption{
    \textbf{Conversion efficiencies of the fundamental spectrum into the harmonics}.
        \textbf{Left:} Integrated intensity contained in the Fundamental spectral domain and the domains of second, third and fourth harmonic generation (SHG THG FHG). The vertical dashed line indicates a maximum on-chip input pulse energy of roughly 60 pJ beyond which the power in the fourth harmonic saturates. 
        \textbf{Right:} Corresponding conversion efficiency.}
  \label{fig:ConversionEfficiency}
\end{figure}

\pagebreak

\section{UV irradiation test of a lithium niobate waveguide}

To assess the susceptibility of our waveguides to UV-induced damage, we performed an additional test, where we coupled a 405~nm laser diode to a lithium niobate waveguide ( 1-1.5~mW of coupled power). To perform this test, we fabricated dedicated lithium niobate waveguides that are not straight but have an S-bend on the waveguide such that input and output are offset from each other (i.e. offset in the direction orthogonal to the direction of the incoming light). This configuration enables accurate measurements of the coupled intensities without picking up stray light that has not propagated through the waveguide. Moreover, a narrow waveguide with a width of 600~nm was chosen to enhance the intensity. Over 12 hours we delivered more than 60~J optical energy through the waveguide, which would correspond to more than one week of UV exposure in astrocomb operation (assuming a UV power level of 100~$\mu$W). We did not observe a noticeable degradation of the UV or infrared transmission through the waveguide within the precision of the measurement. Although the test waveguides were unpoled and would hence not have been suitable for spectrograph calibration, we also performed a supercontinuum generation experiment before and after UV irradiation resulting essentially in the same spectra, as shown in Supplementary Figure~\ref{fig:UV}.

\begin{figure}[ht!]
  \centering
  \includegraphics[width=0.5\textwidth]{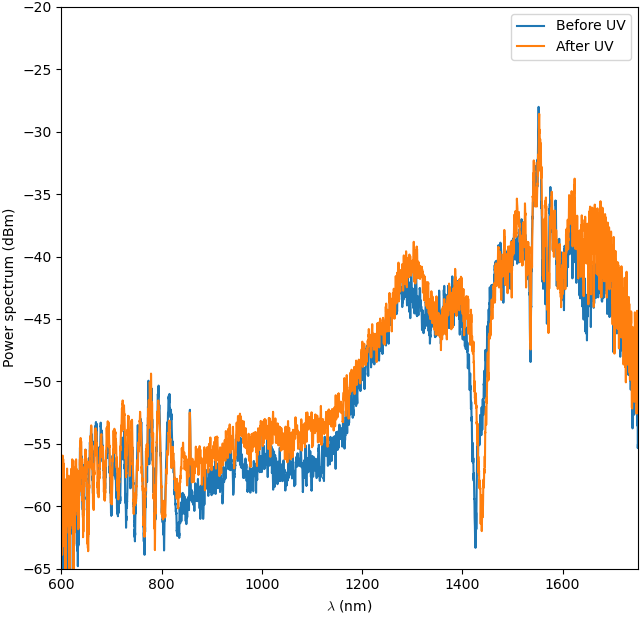}
  \caption{
    \textbf{Supercontinuum before and after UV irradiation.}
        Spectrum of a supercontinuum generated in the lithium niobate waveguide before (blue) and after (orange) UV irradiation. The mode-locked laser driving the supercontinuum has an average power of 73~mW, a pulse duration of 125~fs and repetition rate of 40~MHz.
        }
  \label{fig:UV}

\end{figure}

\section{Broadband spectrum}
\begin{figure}[ht!]
  \centering
  \includegraphics[width=15cm]{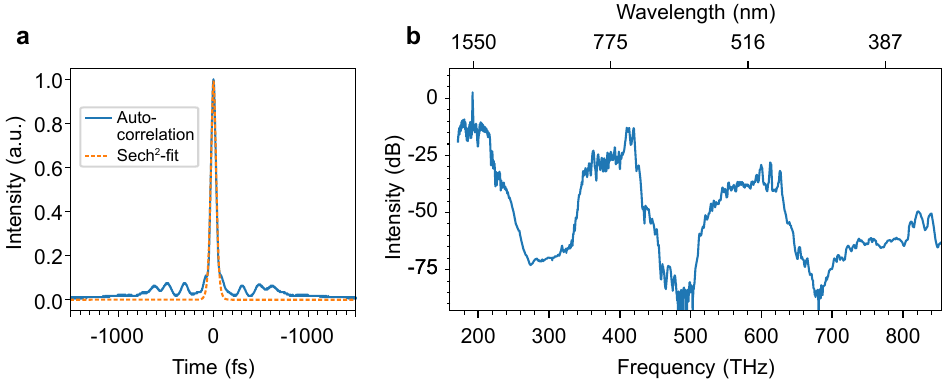}
  \caption{
  \textbf{Broadband spectrum of an 18 GHz comb. a,} Auto-correlation trace and pulse duration based on a $\mathrm{sech^2}$-fit, retrieving a pulse duration of 38~fs (full width at half maximum of the intensity transient). \textbf{b, } Resulting spectrum with overlapping harmonics recorded on a grating based optical spectrum analyzer (OSA, model Yokogawa AQ6374). }
  \label{fig_Broadband}

\end{figure}

To explore the potential of extending the spectral coverage in future work, we configure the 18~GHz electro-optic comb to provide a broadband pulse with an auto-correlation trace as in Figure~\ref{fig_Broadband}a. This is achieved via a combination of highly-nonlinear bandwidth generating fiber and compression fiber, similar to \cite{obrzud2018}. With this input pulse we generate the broadband spectrum of overlapping harmonics as shown in Figure~\ref{fig_Broadband}b.

\printbibliography

\end{refsection}

\end{document}